\input{aipcheck}

\documentclass[
    ,final            
  ]
  {aipproc}

\layoutstyle{6x9}

\begin{document}

\title{Two effects relevant for the study of astrophysical reaction rates: $\gamma$ transitions in capture reactions and Coulomb suppression of the stellar enhancement}

\classification{26.30.-k, 25.40.Lw, 25.40.Hs, 24.60.Dr, 25.20.Dc}
\keywords{reaction rates, nucleosythesis, stellar enhancement, $\gamma$ strength, capture reactions}

\author{Thomas Rauscher}{
  address={Departement Physik, Universit\"at Basel, CH-4056 Basel, Switzerland}
}

\begin{abstract}
Nucleosynthesis processes involve reactions on several thousand nuclei, both close to and far off stability. The preparation of reaction rates to be used in astrophysical investigations requires
experimental and theoretical input. In this context, two interesting aspects are discussed: (i)
the relevant $\gamma$ transition energies in astrophysical capture reactions, and (ii) the newly discovered Coulomb suppression of the stellar enhancement factor. The latter makes a number of
reactions with negative $Q$ value more favorable for experimental investigation
than their inverse reactions, contrary to common belief.
\end{abstract}

\maketitle


\section{Introduction}

Modern nucleosynthesis studies require large reaction networks,
often including thousands of nuclei between the driplines
and their respective reactions
with light particles \cite{cow,schatz,arngor}.
Astrophysical reaction rates employed in reaction network calculations are
determined either directly from cross sections or
from the rate for the inverse reaction by applying detailed balance \cite{adndt}.
The cross sections are either known from experiment or predicted by theory.
Even when a reaction is experimentally accessible,
often astrophysical rates cannot be directly measured.
Excited states are thermally populated in a stellar plasma whereas only reactions on the ground state of the
target can be investigated in the laboratory.
Here I want to address two points which are interesting in the study of astrophysically
relevant reactions, both for experimentalists and for theorists.

\section{Important $\gamma$ transition energies}

\begin{figure}
\includegraphics[width=0.5\textwidth,angle=-90]{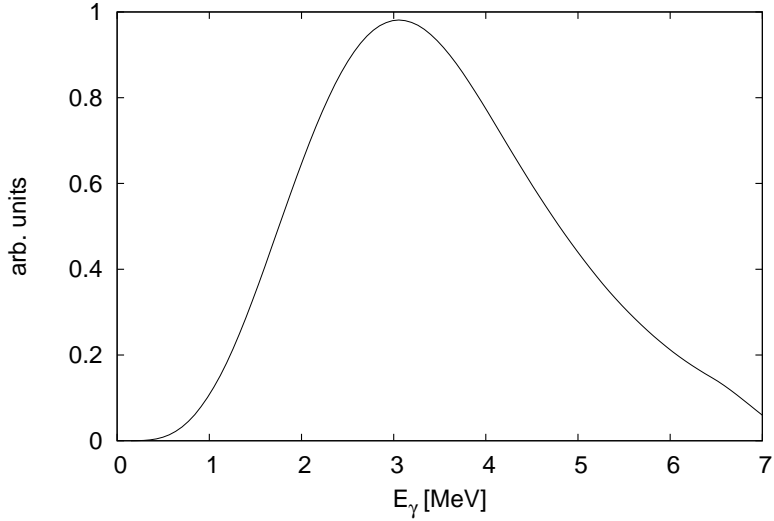}
\caption{Relevant $\gamma$ energy window in $^{124}$Sn for neutron capture on $^{123}$Sn.\label{fig:window}}
\end{figure}

The investigation of electromagnetic transitions at low energies is of high interest to the
fields of both nuclear structure and nuclear reactions. Frequently, the question arises how
astrophysical capture reactions are affected by assumed changes in the photon strength
function (or the nuclear level density). Photon transitions to a state at excitation energy
$E$ considerably increase
in strength with increasing energy $E_\gamma=^\mathrm{target}S_\mathrm{proj}+E_\mathrm{proj}-E$, both for E1 and M1.
On the other hand, the nuclear level density $\rho(E)$ is strongly increasing with increasing $E$, i.e.\
decreasing $E_\gamma$.
Folding the two dependences gives rise to an energy window for the main contributions (see Fig.\ \ref{fig:window}) \cite{raurc}. Fractionation of the smooth behavior may occur at low excitation energy
where only few, isolated excited states are available and the use of a level density function is
not appropriate. However, far off stability there is no information on excited states and most
predictions employ a level density just above the g.s. Sometimes this is, however, appropriate
when excited states are known but transitions to them are suppressed by spin selection rules. Then
transitions in the region of higher level density will still contribute most.

\begin{figure}
\includegraphics[width=0.65\textwidth,angle=-90]{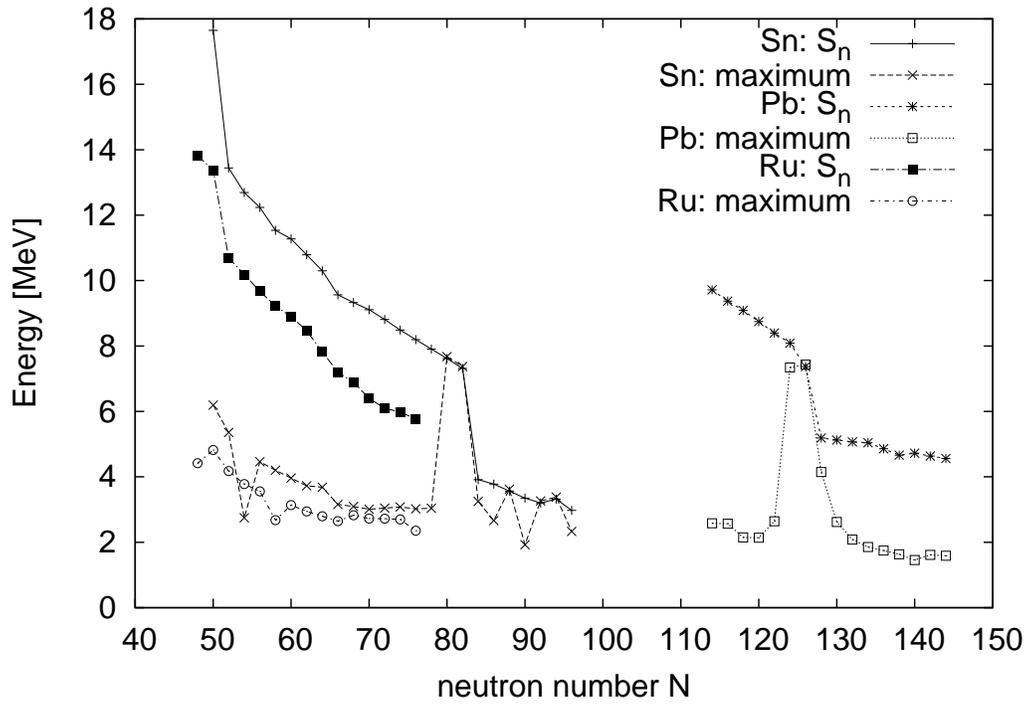}
\caption{Mostly contributing $\gamma$ energy for neutron capture compared to the neutron separation energy in isotopic chains of Ru, Sn, Pb.\label{fig:neutron}}
\end{figure}

\begin{figure}
\includegraphics[width=0.65\textwidth,angle=-90]{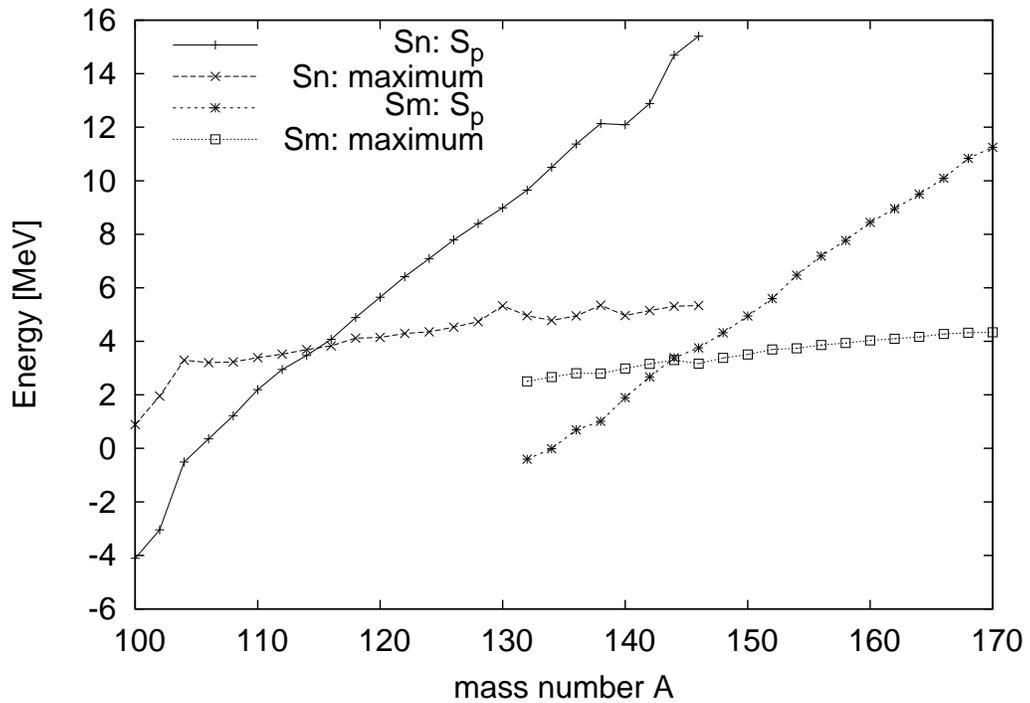}
\caption{Mostly contributing $\gamma$ energy for proton capture compared to the proton separation energy in isotopic chains of Sn and Sm.\label{fig:proton}}
\end{figure}

\begin{figure}
\includegraphics[width=0.65\textwidth,angle=-90]{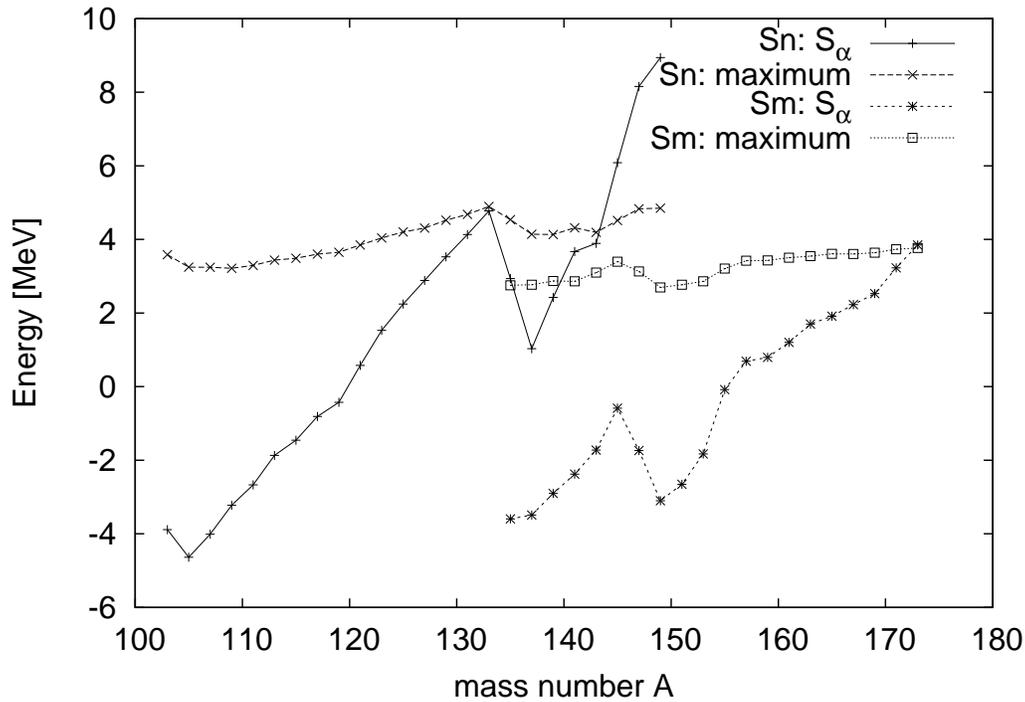}
\caption{Mostly contributing $\gamma$ energy for $\alpha$ capture compared to the $\alpha$ separation energy in isotopic chains of Sn and Sm.\label{fig:alpha}}
\end{figure}

It is instructive to derive the energy window of relevant transitions across the nuclear chart
for neutron, proton, and $\alpha$ capture at astrophysical projectile energies. Figs.\
\ref{fig:neutron}--\ref{fig:alpha}
show the location of the maximum contribution for 60 keV neutrons, 5 MeV protons, and 10 MeV $\alpha$s on isotopic chains of different elements. Due to the higher projectile energies the maximum can lie above the particle separation energy for the charged particle reactions whereas it is always below or at the neutron separation energy for neutron captures. Interestingly, the maximal contribution mostly originates
from $\gamma$ transitions of $2-4$ MeV, quite independent of target and reaction type. Exceptions
are neutron capture reactions to final nuclei with low level density, either at magic numbers or close to the driplines. Because the level density is low and the full energy window cannot be covered by the low-energy neutrons in these cases, the maximal contribution is stemming from transitions close to the
highest possible energy (and allowed by spin selection). An example for this is $^{132}$Sn where
transitions with $E_\gamma=^{^{132}\mathrm{Sn}}S_\mathrm{n}+E_\mathrm{n}\simeq ^{^{132}\mathrm{Sn}}S_\mathrm{n}$ are contributing most. Also neutron captures close to the neutron
dripline exhibit similar behavior. It has to be cautioned, however, that the statistical Hauser-Feshbach model of nuclear reactions is not applicable for nuclei with low level density \cite{rtk97} and
direct capture reactions with $\gamma$ emission to the continuum dominate \cite{raudc}, for which the present considerations are not valid, anyway.

\section{Coulomb suppression of the stellar enhancement}

Astrophysical reaction rates $r^*$ include transitions from thermally excited states $i$ of a nucleus
whereas rates $r^\mathrm{lab}$ derived from laboratory cross sections only consider g.s. transitions.
The importance of excited states is given by the stellar enhancement factor $f=r^*/r^\mathrm{lab}$.
The thermal population follows a Maxwell-Boltzmann statistics $P_i=(2J_i+1)\exp(-E_i/(kT))$.
Additionally, each transition has a transition probability $T_i(J_i,\pi_i,E_i,E_\mathrm{proj},Q)$ so that the total
contribution of transitions from a given state is $C_i=P_iT_i$. It is common to
suggest to measure reactions of astrophysical importance in the direction of positive $Q$ (forward
reaction) because it is generally assumed that $f_\mathrm{forw}<f_\mathrm{rev}$. This is due to the
fact that there are more transitions with higher relative energy $E=|Q|+E_\mathrm{proj}-E_i$ (and thus higher $T_i$) contributing
in the reverse direction.

\begin{figure}
  \includegraphics[width=.65\textwidth,angle=-90]{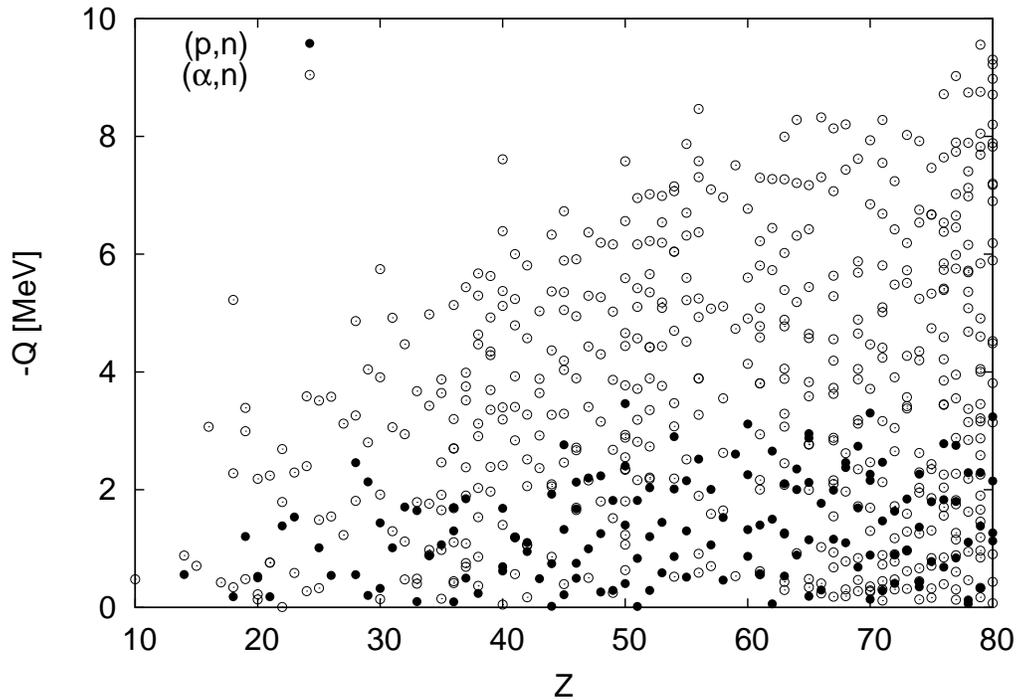}
  \caption{Reaction $Q$ values for (p,n) and ($\alpha$,n) reactions with $1\simeq f_\mathrm{rev}<f_\mathrm{forw}$.\label{fig:qdep}}
\end{figure}

However, reactions having different Coulomb barriers in the entrance and exit channels
may show a converse behavior \cite{kisslett}. If the Coulomb barrier is significantly higher in the reverse direction than in the forward direction, $f_\mathrm{rev}$ may become even smaller than $f_\mathrm{forw}$ because
transitions with slightly lower relative energy are already suppressed (even when possible within spin selection rules) and only the g.s. (or lowest few excited states) contributes. When calculating $f$
for a large range of reactions involving light particles (nucleons, $\alpha$) or photons on targets between the proton- and neutron-driplines more than 1200 cases are
found with $1\simeq f_\mathrm{rev}<f_\mathrm{forw}$. Although there is a dependence on spin (in $P_i$ and selection rules for $T_i$), a dependence on the Coulomb barrier can be clearly seen: larger Coulomb barries allow $f_\mathrm{rev}\simeq 1$ for larger $|Q|$, i.e.\ transitions with larger and larger energies are suppressed. Examples are shown in Fig.\ \ref{fig:qdep} for
(p,n) and ($\alpha$,n) across the nuclear chart. There is a clear correlation between the highest $|Q|$
appearing and the Coulomb barrier (defined by charge $Z$). As expected the dependence on $Z$ is stronger for reactions with the more highly charged $\alpha$ than those involving protons. The scatter for a given
$Z$ is rather determined by the available $Q$ values than by spin effects.

This Coulomb suppression of the stellar enhancement $f$ is important for the conception of
experiments because it implies that it is sometimes better to measure in the direction of negative reaction $Q$ value when being interested in results as close as possible to stellar values. An example is the recently studied reaction $^{85}$Rb(p,n)$^{85}$Sr with $Q=-1.847$ MeV \cite{kisslett,kissthis}.


Support by the Swiss NSF (grant 2000-105328) is acknowledged.


\end{document}